\documentclass[doublecol,a4paper]{epl2} 
\usepackage{amsfonts}
\newcommand{\avg}[1]{\left< #1 \right>} % for average

\title{A random matrix definition of the boson peak}
\shorttitle{A random matrix definition of the boson peak} %Insert here a short version of the title if it exceeds 70 characters

\author{M. L. Manning\inst{1} \and A. J. Liu\inst{2} }
\shortauthor{M. L. Manning  \etal}

\institute{                    
  \inst{1} Department of Physics, Syracuse University, Syracuse, NY 13244 USA\\
  \inst{2} Department of Physics and Astronomy, University of Pennsylvania, Philadelphia, PA 19130 USA
}
\pacs{63.50.-x}{Vibrational states in disordered systems}
\pacs{02.10.Yn}{Matrix theory}
\pacs{71.55.Jv}{Disordered structures; amorphous and glassy solids}
\pacs{62.20.F-}{Deformation and plasticity}

\abstract{The density of vibrational states for glasses and jammed solids exhibits universal features, including an excess of modes above the Debye prediction known as the boson peak located at a frequency $\omega^*$. We show that the eigenvector statistics for boson peak modes are universal, and develop a new definition of the boson peak based on this universality that displays the previously observed characteristic scaling $\omega^* \sim p^{-1/2}$. We identify a large new class of random matrices that obey a generalized global tranlational invariance constraint and demonstrate that members of this class also have a boson peak with precisely the same universal eigenvector statistics.  We denote this class as {\em Boson Peak}  random matrices, and conjecture it comprises a new universality class.  We characterize the eigenvector statistics as a function of coordination number, and find that one member of this new class reproduces the scaling of $\omega^{*}$ with coordination number that is observed near the jamming transition. }

\begin{document}

\maketitle

The normal modes of vibration provide a starting point for understanding the mechanical and thermal response of solids. In both glasses and crystals, the lowest frequency modes are long-wavelength sound modes, and therefore the density of states $D(\omega)$ exhibits Debye scaling, $\omega^{d-1}$, where $d$ is the number of spatial dimensions. In glasses, this scaling is interrupted by a universal band of modes called the boson peak, and these modes play a key role in determining the unique thermal properties of disordered solids~\cite{Phillipsbook,Tanaka,Vitelli2010} and their plasticity~\cite{WidmerCooper, Tanguy1, Manning2011, Chen2011}. 

The boson peak is typically defined as an excess of modes above the Debye prediction~\cite{Phillipsbook}, but this definition is not precise.  While several authors identify the boson peak frequency $\omega^*$ as the value that maximizes $D(\omega)/\omega^{d-1}$~\cite{Grigera, Ciliberti2003},  others identify $\omega^*$ as the frequency at which $D(\omega)$ reaches a quarter or half its maximum value~\cite{Silbert05,Silbert2009,Vitelli2010}.  Unfortunately, these definitions can generate different scalings for $\omega^*$~\cite{SuppFig}.

In addition, while the boson peak consists of a large band of modes, the existing definitions of $\omega^*$ identify only the lowest frequency at which these modes start to appear.  Perhaps the most disturbing aspect of these definitions, however, is that an excess of modes above the Debye prediction is not unique to disordered solids; even perfectly crystalline materials exhibit such an excess~\cite{Ashcroft,Chumakov2011,Chen2013}, typically near a van Hove singularity.  We therefore seek a more robust definition of the boson peak.

The normal modes of vibration of a solid are eigenvectors of the dynamical matrix $M$~\cite{Ashcroft}. For amorphous solids, $M$ is disordered, suggesting that some of its properties can be understood from random matrix theory.  Random matrices have proven useful in many contexts~\cite{Ping1990,Guhr1998} because they yield insight into the most general conditions needed to capture a given behavior.  For example, the eigenvalue spacings for jammed packings of particles are consistent with those of a general class of random matrices, the Gaussian Orthogonal Ensemble ({\em GOE})~\cite{Silbert2009,Zorana}, in which elements are drawn from a Gaussian distribution to form a symmetric matrix~\cite{Mehta}.  More complicated ensembles, including positive definite~\cite{Zippelius,Beltukov2011} and Euclidean Random Matrix ({\em ERM})~\cite{Parisi2002, MezardERM, Amir2010, Amir2013, Goetschy2013} ensembles capture additional universal features, including phonon-like modes at the lowest frequencies and the existence of a boson peak.  {\em ERM} ensembles, based on pairwise interactions between randomly distributed points, also correctly predict that the boson peak shifts to lower frequencies as the solid approaches mechanical instability~\cite{Grigera, Ciliberti2003}.  

All of these results focus on eigenvalue statistics, for good reason.  Until recently, eigenvector statistics for general random matrices have been poorly characterized, except for the {\em GOE}, where they are given by the Porter-Thomas distribution~\cite{Mehta}.  However, mathematicians have now shown that the {\em GOE} is part of a large class of matrices with ``maximally delocalized," eigenvectors, where no eigenvector entry is larger than $O(\sqrt{N})$ and $N$ is the linear size of the matrix~\cite{Erdos2009,TaoReview}. In addition, a few studies~\cite{Barbosa2000, Hamoudi2002, Peplowski1993} have characterized eigenvector statistics for more complicated matrices using fitting functions or superpositions of Porter-Thomas distributions, although they did not observe universality. 

In this Letter, we use random matrices to provide a more robust definition of the boson peak in disordered solids, identify a set of minimal requirements necessary to generate this feature, and describe how the boson peak changes with coordination number.  We first show that eigenmodes in the boson peak of jammed solids have a universal structure, and that the onset of universality occurs at a frequency $\omega^{*}$ that scales with the pressure $p^{1/2}$, precisely as predicted by marginal stability arguments for the boson peak~\cite{Wyart2005,arcmp}.  This suggests a new definition of the boson peak based not on comparison of the vibrational spectrum to Debye scaling, but rather on the eigenvector statistics.    
\begin{figure}[h!]
\centering \includegraphics[width=0.45\textwidth]{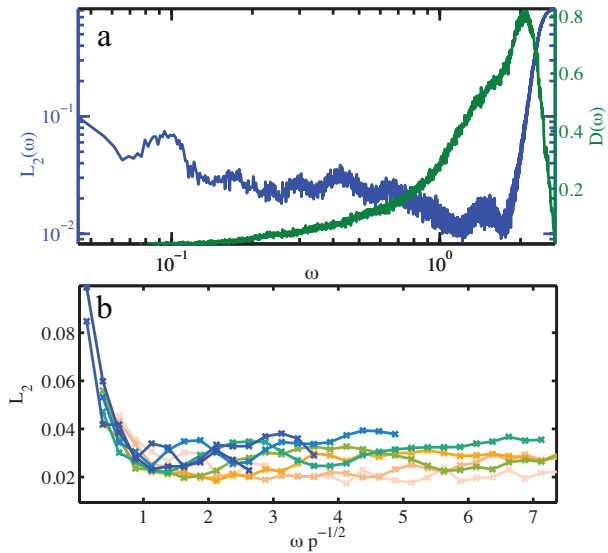}
\caption{\label{BPev2}{\bf Universality of eigenstatistics for jammed packings}  {\bf (a)} Eigenstatistics as a function of eigenvalue frequency $\omega$ for a jammed packing at a pressure of $p=10^{-1.4}$.  The green line (right axis) indicates the the density of states $D(\omega)$.  The solid line (left axis) is the $L_2$ difference between $P(\tilde{ \ell})|_{\omega}$ and the universal distribution $P_{GW}(\tilde{ \ell})$ (black dashed line in panel (a)). {\bf (b)} The $L_2$ difference between $P(\tilde{ \ell})$ and $P_{GW}(\tilde{ \ell})$ collapses as a function of the rescaled variable $\omega p^{-1/2}$. Different lines correspond to different pressures logarithmically spaced from $p=10^{-1.4}$ (dark blue line) to $p=10^{-4.2}$ (light pink line). }
\end{figure}

In addition, we identify a large class of dense random matrices, which includes {\em ERM}     as a subset, that all possess a band of eigenvectors with this same universal structure.  We demonstrate that the eigenvector statistics for this class are \emph{not maximally delocalized}, so that this class of matrices is distinct from the previously identified ensemble of matrices with maximally delocalized eigenvectors~\cite{Erdos2009,TaoReview}.  We conjecture that these matrices comprise a new universality class that we call the Boson Peak ({\em BP}) ensemble.  To understand the role of coordination number, we also study a class of sparse matrices -- the Sparse Boson Peak ({\em SBP}) ensemble--and demonstrate that their eigenvector statistics converge rapidly to the universal {\em BP} distribution as the number of non-zero entries increases. Finally, we demonstrate that one member (diagonally dominant or DD matrices) of the {\em SBP} ensemble reproduces the scaling of the boson peak frequency with coordination number seen in jammed packings. Taken together, these results indicate that the boson peak is generated by the interplay between disorder and translational invariance, and provide a simple explanation for the universality of the boson peak in glasses.
\begin{figure}[h!]
\centering \includegraphics[width=0.45\textwidth]{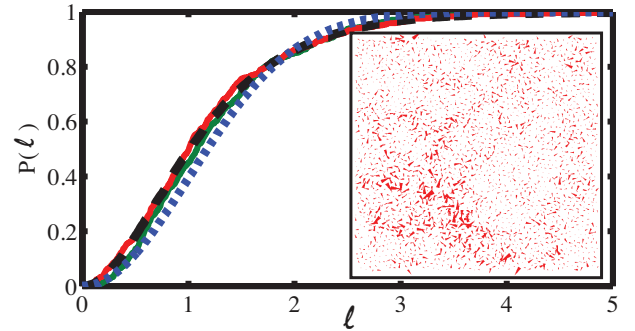}
\caption{\label{BPev1} {\bf Cumulative distribution functions for eigenvector statistics.} Solid lines are for individual vibrational modes of a jammed packing at $z=4.3$ at two different frequencies above $\omega^* = 1.16$: $\omega = 1.17 (red), 1.28 (green)$, in the boson peak of a model 2D jammed solid. Dashed line is for the {\em GW} ensemble  ($P_{GW}(\tilde{ \ell})$) defined in the main text, and dotted blue line is the analytic result for the Gaussian Orthogonal Ensemble ({\em GOE}). (Inset)  A real-space representation of a vibrational mode from the boson peak in a jammed solid. }
\end{figure}

Our starting point is the dynamical matrices of jammed packings.  Each eigenvalue $E$ corresponds to a vibrational frequency $\omega \sim \sqrt{E}$.  For solids, the density of vibrational states $D(\omega)$ follows the Debye scaling $\omega^{d-1}$ expected for acoustic modes at the lowest frequencies.  Above a characteristic boson peak frequency $\omega^*$, the spectrum no longer obeys  Debye scaling. Below $\omega^{*}$, there is a population of modes that are quasi localized hybridizations of localized excitations with extended phonon-like modes~\cite{Zorana,Xu2010}. Just above $\omega^{*}$ the modes are extended and disordered, and at the highest frequencies near the band edge they are completely localized~\cite{Zorana,Silbert2009}.  The green line (right axis) in Fig~\ref{BPev2}(a) is a plot of the density of states $D(\omega)$, with $\omega^{*} \sim 1.16$, averaged over 500 simulated particle packings at a pressure of $10^{-1.4} \sim 0.04$.  Each packing consists of 512 mechanically stable bidisperse harmonic soft disks in 2D~\cite{OhernCG}, with eigenstatistics calculated by diagonalizing the dynamical matrix of linear size $N = 2\times 512= 1024$. Lengths are in units of the average particle diameter and energies are in units of the harmonic spring constant.

The inset to Fig~\ref{BPev1} illustrates the spatial organization of a typical vibrational mode near $\omega^*$. This mode is extended and heterogeneous with vortex-like features predicted by a variational argument near isostaticity~\cite{WyartBrito}.  The spatial pattern of each mode is defined by  the set $\{ \ell_{i,\alpha}\}$,  where $ \ell_{i,\alpha}$ is the magnitude of the polarization vector (the displacement of particle $\alpha$ in eigenvector $i$) and varies drastically from mode to mode in the boson peak. 

Remarkably, however, the probability distribution function for the magnitudes $\ell_\alpha$ is nearly identical for different modes in the boson peak. We focus on the cumulative distribution function (cdf) $P_{BP}(\tilde{\ell})$, where $\tilde{ \ell}$ is the normalized magnitude $ \ell \sqrt{N}$. The solid lines in Fig~\ref{BPev1} illustrate the cumulative distribution for several individual modes near $\omega^*$.  

To quantify and study this universality, we first identify a universal cdf $P_{GW}(\tilde{ \ell})$ that closely approximates eigenmodes with frequencies near $\omega^*$, shown by the black dashed line in Fig~\ref{BPev1}.  Rather than averaging over many boson peak modes, we generate this distribution from an ensemble of simple random matrices, as discussed in detail below. For the remainder of this Letter, all cumulative distributions shown are for matrices of linear size $N=1024$ (jammed model solid) or $N=1000$(random matrices), averaged over $20$ modes in the same frequency window to reduce numerical fluctuations due to finite system size. 

 The difference between each boson peak cdf $P_{BP}(\tilde{\ell})$ and the universal distribution $P_{GW}(\tilde{ \ell})$ is quantified using the $L_2$ norm $\int (P_{BP}-P_{GW})^2 dl$. The blue line (left axis) in Fig~\ref{BPev2}(a) shows this $L_2$ norm as a function of frequency. Phonon-like modes at low frequencies and Anderson-localized modes at high frequencies are very different from the universal distribution and have large $L_2$ differences.  In contrast, there is a broad minimum in the $L_2$ norm near $\omega^{*} \sim 1.16$, corresponding to a large band of modes with the same universal structure near the boson peak.  Fig~\ref{BPev2}(b) is a plot of the low-frequency behavior of this $L_2$-norm for packings at different pressures $p$.  The location of the minimum collapses when $\omega$ is rescaled by $p^{1/2}$, indicating that onset of universality in eigenmodes (the minimum in the L2 norm) occurs at a frequency $\omega^* \sim p^{1/2}$.  This is the scaling associated with the boson peak for harmonic disks~\cite{Silbert05,Wyart2005,arcmp}.   Our results therefore suggest a new definition of the boson peak frequency based on eigenvector statistics: the boson peak frequency is the frequency at which the polarization vector magnitude distribution is closest to the universal distribution $P_{GW}$, so that the $L_2$ difference between the two distributions is at a minimum.
 
\begin{figure}[h!]
\centering \includegraphics[width=0.45\textwidth]{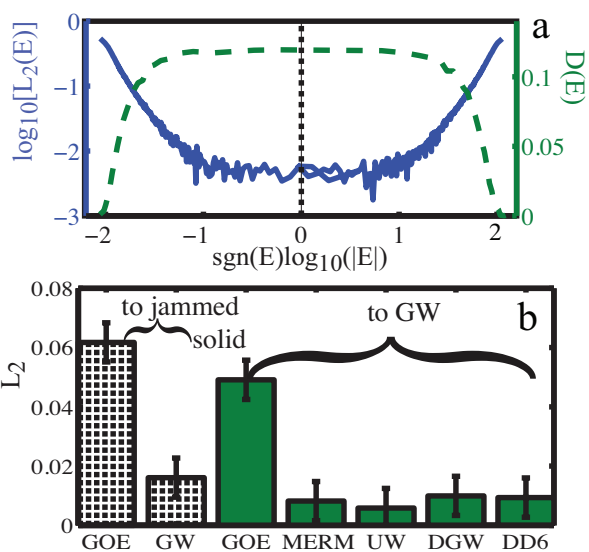}
\caption{\label{GWev} {\bf (a)} Eigenstatistics as a function of eigenvalue energy $E$ for the {\em GW} ensemble.  The dotted line indicates the peak in the density of states at $\bar{E}=0$.  The solid line (left axis) is the $L_2$ difference between $P(\tilde{ \ell})|_E$ and $P(\tilde{ \ell})|_{\bar{E}=0}$. The dashed line (right axis) is the density of states. {\bf (b)} $L_2$ distance between eigenvector statistics $P(\tilde{ \ell})|_{\bar{E}}$ for different random matrix ensembles, as defined in the text. DD6 is the diagonally dominant ensemble with ${\mathfrak z} =6$. Hatched bars compare boson peak modes in jammed packings to random matrix eigenvectors, solid bars compare  {\em GW} to other ensembles. Error bars indicate the average $L_2$ difference between two members of the same ensemble.}
\end{figure}

The obvious remaining question is the origin of the observed universality.  To understand the surprising similarity in the eigenvector statistics for these modes, we turn to random matrices. We first compare the boson peak eigenvector statistics to those for the {\em GOE}. Because each displacement magnitude in a boson peak mode from a 2D jammed solid is the vector sum of two mode entries, we use $P_{GOE}(\tilde{ \ell}) = (1 - \exp(-\tilde{ \ell }^2/2)) $, the Porter-Thomas distribution for a {\em GOE} eigenvector $\{ \ell_{\alpha}\}$ where pairs of components $\ell_{\alpha}, \ell_{\alpha+1}$ have been vector summed: $\tilde{ \ell_{\tilde{\alpha}}} = \sqrt(\ell_{\alpha}^2 + \ell_{\alpha+1}^2)$. This is shown as the blue dotted line in Fig.~\ref{BPev1}.  While it is close to the distributions for the boson peak modes, there are clearly systematic deviations. The difference between the boson peak and the {\em GOE} cdf distributions is quantified using the $L_2$ norm $\int (P_{BP}-P_{GOE})^2 dl$,  and corresponds to the first hatched box in Fig~\ref{GWev}(b). Thus, although the GOE results are consistent with the \emph{eigenvalue} statistics of the boson peak~\cite{Silbert2009,Zorana}, the GOE fails to capture the \emph{eigenvector} statistics.

One obvious difference between {\em GOE} matrices and dynamical matrices lies in the statistics of the on-diagonal elements.  Because the potential energy of jammed packings only depends on differences in the displacements of interacting particles, the dynamical matrix is translation invariant. As a result, diagonal elements, which correspond to self-interaction terms, obey a sum rule: $ M_{ii} = - \sum_{j \neq i} M_{ij}$.   The sum rule can be imposed on a random matrix, as it is for Euclidean random matrices.  Here, we explore whether the sum rule is essential by investigating a class of symmetric Wigner matrices~\cite{TaoReview}, which do not obey the sum rule, but whose diagonal elements have similar statistics.  We will refer to this ensemble as the Gaussian Wigner ({\em GW}) ensemble.

Specifically, the  {\em GW} ensemble consists of symmetric matrices with linear size $N$, where the off-diagonal elements are independent random variables taken from a Gaussian distribution with mean $\mu$ and variance $\sigma^2$, and the on-diagonal elements are independent and Gaussian distributed with mean $-N \mu$ and variance $N \sigma^2$.  We refer to this condition on the on-diagonal elements as a generalized translational invariance constraint.

  The dashed line in Fig~\ref{GWev}(a) shows the density of eigenvalues $D(E)$ for the {\em GW}.  The peak in the density of states occurs at $\bar{E}=0$ and the cdf corresponding to that eigenvalue, $P_{GW}(\tilde{ \ell}) |_{\bar{E}=0}$, is the black dashed line in Fig.~\ref{BPev1}. In contrast to the {\em GOE}, eigenvectors in the {\em GW} ensemble vary with the eigenvalues $\{ E \}$.   The solid line in Fig~\ref{GWev}(a) compares the cdf for eigenvectors at various values of $E$ to $P_{GW}(\tilde{ \ell}) |_{\bar{E}=0}$. The $L_2$ difference is large where $D(E)$ falls off sharply, indicating modes that are highly localized at the band edge and different in character from those at $\bar{E}$.

However, there is a well-defined band of modes in the middle of the spectrum (near $\bar{E}$) that each possess the same universal eigenvector distribution. We find that this distribution $P_{GW}( \tilde{ \ell})|_{\bar{E}} \equiv P_{GW}(\tilde{ \ell})$ does not depend on our choice for $\mu$ or $\sigma$ or the linear matrix size $N$.  Fluctuations from eigenvector to eigenvector are quantified by the $L_2$-distance between cumulative distributions for an individual eigenvector and for the average over 20 eigenvectors centered around $\bar{E}$; these decrease approximately as $1/\sqrt{N}$.  Thus, each individual eigenvector in the middle of the spectrum approaches the same well-defined limiting distribution, which we call the {\em BP} distribution, in the thermodynamic limit.  Finally, we compare the {\em BP} distribution to the cumulative eigenvector distribution for modes at the boson peak of jammed packings. The second hatched box in Fig.~\ref{GWev}(b) shows that the two distributions are nearly identical, suggesting that the modes of the boson peak in jammed packings are very similar to eigenvectors near $\bar{E}$ in the GW ensemble.  This validates using $P_{GW}(\tilde{ \ell})$ to approximate the universal distribution for jammed packings in Fig.~\ref{BPev1}.

The fact that the eigenvector statistics for the {\em GW} ensemble are distinct from those of the {\em GOE} is surprising, because it has been proven that a large class of Wigner matrices have maximally delocalized eigenvectors, just like the {\em GOE}~\cite{Erdos2009,TaoReview}.  Our {\em GW} ensemble does not satisfy the assumptions of the proof because the variance of the on-diagonal elements grows linearly with the system size $N$ and is therefore unbounded in the thermodynamic limit.

%This linear dependence on $N$ appears to be critical: ensembles where on-diagonal variance scales with different powers of $N$ have different eigenvector statistics that do not fall into our universality class. 

%
\begin{figure}[h!]
\centering \includegraphics[width=0.45\textwidth]{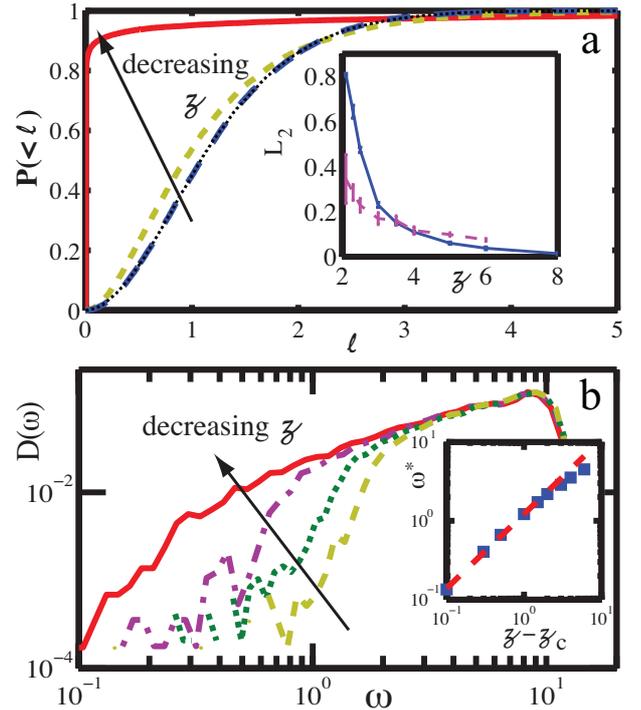}
\caption{\label{coord1} {\bf (a)} Eigenvector statistics $P(\tilde{ \ell})$ for sparse DD matrices at different values for the average coordination number ${\mathfrak z}$ (dotted  = 1000, long dash=6, short dash = 3.5, solid = 2.1). (inset) $L_2$ distance as a function of ${\mathfrak z}\!-\!{\mathfrak z}_c$ for SGW (dashed line) and DD (solid line) matrices.  {\bf (b)}  Density of states for diagonally dominant (DD) random matrix ensembles captures the shift in the boson peak location to lower frequencies as average coordination number ${\mathfrak z}$ decreases (short dash = 3.5, dotted = 3, dash-dot=2.5, solid = 2.1).(inset) Blue squares indicate the frequency of the boson peak $\omega^{*}$ as a function of $\delta {\mathfrak z}  = {\mathfrak z}-{\mathfrak z}_c$. Red dashed line is best fit to lowest 5 points with slope $0.94$. }
\end{figure}

 A natural question is whether these results depend on the underlying distribution for the random matrix entries. Therefore, we study other Wigner matrix ensembles where off-diagonal entries are chosen from a distribution $p_{\mu, \sigma}(x)$ with finite first ($\mu$) and second ($\sigma$) moments, while the on-diagonal elements are Gaussian distributed random variables with mean $-N \mu$ and variance $N \sigma^2$.   We denote matrices with $p(x) = 1/d \in [0,d] $ as Uniform Wigner ({\em UW}),  and those with $p(x) = A \exp( (x- \mu_1)/\sigma_1^2 +  B \exp( (x- \mu_2)/\sigma_2^2 $ as Double Gaussian Wigner ({\em DGW}). As shown by Fig~\ref{GWev}(b), both of these matrix ensembles are indistinguishable from  {\em GW} and belong in our proposed universality class. For $N=1000$ their difference is $L_2=0.0058$ and $L_2 = 0.0099$, respectively, which are the same order as the average $L_2$ difference between distributions for different members of $GW$ ensemble.

 We expect that random matrices constructed in this manner from any continuous distribution $p(x)$ with finite first and second moments will be characterized by having the same eigenvector distribution function.  Thus, we conjecture that these matrices belong to a universality class, which we will call the Boson Peak ({\em BP}) ensemble, characterized by the universal {\em BP} eigenvector distribution at $\bar{E}$, the peak in their density of states.

We note that $\bar{E}$ is different from the boson peak frequency $\omega^*$.  While $\omega^*$ is typically identified as the lowest frequency at which a solid's vibrational spectrum deviates from Debye scaling -- i.e. the low-frequency edge of a band of boson peak modes -- $\bar{E}$ is the \emph{average} vibrational energy for that band of modes.  In fact, we find that over 30 \% of the vibrational modes in jammed packings have statistics that are indistinguishable ($L_2 < 0.02$) from the universal {\em GW} distribution, quantifying our intuition that there is very large band of extended, disordered BP modes that begins at $\omega^*$ and extends far into the spectrum.
  
We now consider a class of matrices that explicitly obey the sum rule required for translation invariance, so that the off-diagonal elements are Gaussian distributed and the on-diagonal elements are the negative sum of the off-diagonal elements in each row.   We denote these as {\em MERM}, because they are the mean-field (or infinite-ranged) limit of a Euclidian random matrix ensemble with Gaussian interaction potentials~\cite{MezardERM}.  The sum rule leads to strong correlations between matrix elements within a row.  As a result, these ensembles exhibit one zero-energy eigenvector with identical entries. This corresponds to a uniform displacement associated with global translation invariance. Eigenvector statistics for the remaining $N-1$ modes are indistinguishable from those in the  {\em GW} ensemble, as shown in Fig.~\ref{GWev}(b), and so  {\em MERM} matrices belong to the proposed BP universality class.  

So far, we have considered dense random matrix ensembles where all off-diagonal elements are drawn from a continuous distribution.  In contrast, the dynamical matrix of a jammed packing is sparse, with an average of only ${\mathfrak z} = z d$ nonzero off-diagonal elements in each row, where $d$ is the dimensionality and $z$ is the number of interacting neighbors per particle.  We therefore study two ensembles of sparse random matrices.  

The first sparse ensemble possesses ${\mathfrak z}$ nonzero off-diagonal elements per row chosen from the distributions characterizing the  {\em GW} ensemble and is denoted  {\em SGW}.    As shown by the dashed line in the inset to Fig.~\ref{coord1}(a), the eigenvector statistics for sparse matrices depend on ${\mathfrak z}$. However, the distributions converge rapidly; those for ${\mathfrak z}\! = \!  6$ and ${\mathfrak z} = N \! = \!1000$ are nearly indistinguishable.  This explains why the boson peak modes for 2D jammed packings (with $\avg{ z }=5.1$ and $\avg{  {\mathfrak z} } = 10.2$) are so close to the universal distribution with ${\mathfrak z}=N=1000$, as shown by the second hatched box in Fig~\ref{GWev}(b).   The modes of the boson peak of jammed packings approach the universal BP distribution in the large ${\mathfrak z}$ limit.

To ensure mechanical stability, the dynamical matrices for jammed packings are positive definite, but the random matrix ensembles we have considered so far are not.  This demonstrates that the eigenvector statistics -- i.e. the boson peak mode structure -- do not depend on mechanical stability.  However, none of these random matrices captures the location of the boson peak (at an eigenvalue $E^*$ and frequency $\omega^* = \sqrt{E^*}$) that is intimately related to mechanical stability~\cite{Ciliberti2003, Silbert05, Wyart2005, Wyart2010}. 

To capture the boson peak frequency, we turn to a sparse positive definite ensemble known as symmetric Diagonally Dominant ({\em DD}) matrices, which we generate as follows.  First, we populate an integer number $\zeta_i$ of off-diagonal entries in each row $i$ such that there is an average $\avg{ \zeta_i} =  {\mathfrak z}/2$ entries in each row.  These nonzero entries are random variables chosen from a uniform distribution on the interval $[ -1, 0 ]$. To obtain a symmetric matrix, this matrix is added to its transpose so that the average coordination number is $\mathfrak z$.  Finally, the diagonal elements are the negative sum of the off-diagonal entries in the same row.  One can show that random matrices constructed in this manner are always positive definite~\cite{HornMatrix}, and therefore each eigenvalue $E$ corresponds to a frequency: $\omega = \sqrt{E}$.  Just as in the {\em MERM} ensemble, we have introduced strong correlations within a row to maintain positive definiteness. 

We study the behavior of the eigenvector statistics and density of states for {\em DD} matrices as a function of their average coordination number.  As shown by Fig.~\ref{coord1}(a) the eigenvector statistics quickly approach the universal BP distribution at large ${\mathfrak z}$.  Therefore dense {\em DD} matrices belong to the BP universality class.  To compare to the dynamical matrix for jammed packings, we note that in the latter systems a sum rule must be enforced separately in each direction, which means that each row contains $d$ subblocks (one per dimension), and the sum rule relating the diagonal elements to the sum of off-diagonal elements must be obeyed within each subblock.  Because the {\em DD} matrices in our proposed universality class obey only one sum rule, they are effectively one-dimensional and the average number of nonzero off-diagonal elements in a row, ${\mathfrak z}$, is the same as the number of neighbors per particle, $z$.  Thus, the critical coordination number for isostaticity is ${\mathfrak z}_c \!=\!z_c  \!=\!2d \!=\!2$. 

Note that the $L_2$-distance between the $DD$ eigenvector statistics and the BP distribution increases rapidly as ${\mathfrak z} \rightarrow 2$.   Thus, the BP distribution fails to describe modes in the double limit $\omega \rightarrow \omega^*$ and ${\mathfrak z} \rightarrow 2$ in this one-dimensional case.  However, Fig~\ref{BPev2}(b) shows that the $BP$ ensemble provides an excellent description in the same double limit in two dimensions (where $z_c=4, {\mathfrak z}_c =8$) since packings at low $p$ have values of $z$ near $z_c$.  The large values of $L_2$ near ${\mathfrak z} \!=\!2$ therefore appear to result from the fact that ${\mathfrak z} \!=\!2$  is simply too far away from the mean-field, large ${\mathfrak z}$ limit.

In the {\em DD} ensemble the density of modes $D(\omega)$ drops off rapidly below a characteristic frequency $\omega^*$ that depends on the coordination ${\mathfrak z}$, as shown in Fig.~\ref{coord1}(b). The inset to Fig.~\ref{coord1}(b) demonstrates that $\omega^*$ (defined as the lowest frequency at which $D(\omega)$ attains half of its maximum value) scales linearly with $ \delta {\mathfrak z}  = {\mathfrak z}  - {\mathfrak z}_c$. This is precisely the scaling seen in jammed packings~\cite{Silbert05}.  

In summary, our results demonstrate that dense random matrices of linear size $N$, where the on-diagonal elements have a variance $N$ times larger than the off-diagonal elements, possess a universal eigenvector structure at the peak in their spectrum, the BP distribution, that is distinct from the standard {\em GOE} Porter-Thomas distribution.  Furthermore, sparse random matrices with coordination $\mathfrak z$, where on-diagonal elements have a variance $\mathfrak z$ times larger than the off-diagonal elements, rapidly approach this universal distribution as $\mathfrak z$ increases.  For $d \ge 2$, the boson peak modes are well-described by the BP distribution and are therefore a generic result of disorder and global translation invariance.

Interestingly, matrices with entries from discrete probability distributions do not appear to be in the same universality class -- Zippelius and co-workers~\cite{Zippelius} have shown that adjacency matrices (where off-diagonal elements are either one or zero) possess discrete jumps in the density of states with corresponding localized eigenvectors. 

In addition, our results demonstrate that at frequencies below $\omega^{*}$, eigenvectors begin to depart from universal behavior, and modes corresponding to structural flow defects begin to appear~\cite{Manning2011}. Finally, our results corroborate previous work demonstrating that the frequency at which the boson peak occurs is closely related to mechanical stability.  By decreasing the coordination number towards isostaticity in our {\em DD} ensembles, we recover the characteristic scaling of the boson peak frequency with contact number seen in jammed packings.

Many real glassy solids interact via long-ranged potentials.  As a result, their dynamical matrices are more dense than those of jammed packings but still obey the sum rule.  Network glasses or metallic glasses likewise must obey the sum rule and yet are disordered.  Our random matrix analysis indicates that the dynamical matrices of all these systems should belong to the BP ensemble so all of these disordered solids should exhibit boson peaks--a band of modes characterized by the BP distribution.  %Indeed, several such systems are well known to exhibit an excess of modes relative to the Debye prediction~\cite{Phillipsbook}.  

Thus, our analysis suggests an explanation for the existence of boson peak modes in glassy solids that is complementary to that of Wyart, et al.~\cite{Wyart2005}.  We expect to see a band of boson peak modes in any disordered material for the same reason we expect to see a universal distribution of mode frequencies  -- it is a necessary consequence of the interplay between disorder and translational invariance.  The frequency of the boson peak, on the other hand, is much more sensitive to isostaticity and sparseness of the random matrix.
%\begin{figure}
%\onefigure{epl-template.eps}
%\caption{Figure caption.}
%\label{fig.1}
%\end{figure}

%\emph{Acknowledgements}
\acknowledgements We thank C. P. Goodrich, N. Cook, T. Tao, and W. E. Ellenbroek for discussions, and C.P.G. for simulation data. This research was supported by the U.S. DOE Award DE-FG02-05ER46199 (AJL) and by NSF-DMR-1352184 (MLM).

%\begin{thebibliography}{0}

%\bibitem{b.a}
 % \Name{Author F., Author S. \and Author T.}
 % \REVIEW{Some Rev. A}{69}{1969}{9691}.


\begin{thebibliography}{10}
\expandafter\ifx\csname url\endcsname\relax\def\url#1{\texttt{#1}}\fi

\bibitem{Phillipsbook}
\Name{Phillips W.} (Editor) \Book{Amorphous Solids: Low-Temperature Properties}
(Springer-Verlag Berlin) 1981.

\bibitem{Tanaka}
\Name{Shintani H. \and Tanaka H.} \REVIEW{Nat. Mat.}{7}{2008}{870}.

\bibitem{Vitelli2010}
\Name{Vitelli V. \emph{et al.}} \REVIEW{Phys.
  Rev. E}{81}{2010}{021301}.

\bibitem{WidmerCooper}
\Name{Widmer-Cooper A. \and Harrowell P.} \REVIEW{Phys. Rev.
  Lett.}{96}{2006}{185701}.

\bibitem{Tanguy1}
\Name{Tsamados M. \emph{et al.} } \REVIEW{Phys.
  Rev. E}{80}{2009}{026112}.

\bibitem{Manning2011}
\Name{Manning M.~L. \and Liu A.~J.} \REVIEW{Phys. Rev. Lett.}{107}{2011}{108302}.

\bibitem{Chen2011}
\Name{Chen K. \emph{et al.}} \REVIEW{Phys. Rev. Lett.}{107}{2011}{108301}.

\bibitem{Grigera}
\Name{Grigera T.\emph{et al.}}
  \REVIEW{J. Phys.: Cond. Mat.}{14}{2002}{2167}.

\bibitem{Ciliberti2003}
\Name{Ciliberti S. \emph{et al.} }
  \REVIEW{J. Chem. Phys.}{119}{2003}{8577}.

\bibitem{Silbert2009}
\Name{Silbert L.~E., Liu A.~J. \and Nagel S.~R.} \REVIEW{Phys. Rev.
  E}{79}{2009}{021308}.

\bibitem{SuppFig}
\Book{Supp. Figs: https://mmanning.expressions.syr.edu/epl2015/}

\bibitem{Ashcroft}
\Name{Ashcroft N. \and Mermin N.} \Book{Solid state physics}
  (Holt, Rinehart and Winston) 1976.

\bibitem{Chumakov2011}
\Name{Chumakov A.~I. \emph{et al.} }
  \REVIEW{Phys. Rev. Lett.}{106}{2011}{225501}.

\bibitem{Chen2013}
\Name{Chen K. \emph{et al.}} \REVIEW{Phys. Rev. E}{88}{2013}{022315}.

\bibitem{Ping1990}
\Name{Ping S.} \Book{Scattering and Lozalization of Classical Waves in Random
  Media} Vol.~8 (World Scientific) 1990.

\bibitem{Guhr1998}
\Name{Guhr T., M{\"u}ller-Groeling A. \and Weidenm{\"u}ller H.~A.}
  \REVIEW{Phys. Rep.}{299}{1998}{189}.

\bibitem{Zorana}
\Name{Zeravcic Z., van Saarloos W. \and Nelson D.~R.} \REVIEW{Europhys.
  Lett.)}{83}{2008}{44001}.

\bibitem{Mehta}
\Name{Mehta M.~L.} \Book{Random matrices} Vol. 142
  2004.

\bibitem{Zippelius}
\Name{Broderix K. \emph{et al.}}
  \REVIEW{Phys. Rev. E}{64}{2001}{021404}.

\bibitem{Beltukov2011}
\Name{Beltukov Y. \and Parshin D.} \REVIEW{Physics of the Solid
  State}{53}{2011}{151}.

\bibitem{Parisi2002}
\Name{Parisi G.} \REVIEW{Euro. Phys. J. E}{9}{2002}{213}.

\bibitem{MezardERM}
\Name{M{\'e}zard M., Parisi G. \and Zee A.} \REVIEW{Nuc. Phys.
  B}{559}{1999}{689}.

\bibitem{Amir2010}
\Name{Amir A., Oreg Y. \and Imry Y.} \REVIEW{Phys. Rev.
  Lett.}{105}{2010}{070601}.

\bibitem{Amir2013}
\Name{Amir A. \emph{et al.}} \REVIEW{Phys.
  Rev. X}{3}{2013}{021017}.

\bibitem{Goetschy2013}
\Name{Goetschy A. \and Skipetrov S.} \REVIEW{arXiv preprint 1303.2880}{}{2013}{}.

\bibitem{Erdos2009}
\Name{Erd{\H{o}}s L., Schlein B. \and Yau H.-T.} \REVIEW{Ann.
  Prob.}{}{2009}{815}.

\bibitem{TaoReview}
\Name{Tao T. \and Vu V.} \REVIEW{arXiv preprint 1202.0068}{}{2012}{}.

\bibitem{Barbosa2000}
\Name{Barbosa C., Guhr T. \and Harney H.} \REVIEW{Phys. Rev.
  E}{62}{2000}{1936}.

\bibitem{Hamoudi2002}
\Name{Hamoudi A. \emph{et al.}}
  \REVIEW{Phys. Rev. C}{65}{2002}{064311}.

\bibitem{Peplowski1993}
\Name{Peplowski P. \and Haake F.} \REVIEW{J. Phys. A}{26}{1993}{3473}.

\bibitem{Wyart2005}
\Name{Wyart M., Nagel S. \and Witten T.} \REVIEW{Europhys.
  Lett.}{72}{2005}{486}.

\bibitem{arcmp}
\Name{Liu A.~J. \and Nagel S.~R.} \REVIEW{Annu. Rev. Condens. Matter
  Phys.}{1}{2010}{347}.

\bibitem{Xu2010}
\Name{Xu N. \emph{et al.}} \REVIEW{Europhys.
  Lett.}{90}{2010}{56001}.

\bibitem{OhernCG}
\Name{O'Hern C.\emph{et al.}} \REVIEW{Phys. Rev.
  Lett.}{88}{2002}{075507}.

\bibitem{WyartBrito}
\Name{Brito C. \and Wyart M.} \REVIEW{J. Stat. Mech.}{2007}{2007}{L08003}.

\bibitem{Silbert05}
\Name{Silbert L., Liu A. \and Nagel S.} \REVIEW{Phys. Rev.
  Lett.}{95}{2005}{098301}.

\bibitem{Wyart2010}
\Name{Wyart M.} \REVIEW{Europhys.
  Lett.}{89}{2010}{64001}.

\bibitem{HornMatrix}
\Name{Horn R. \and Johnson C.} \Book{Matrix Analysis} (Cambridge University
  Press) 1990.

\end{thebibliography}
\end{document}